\documentclass[pdftex]{article}
\usepackage{icrctc07}

\title{Schwarzschild-Couder two-mirror telescope for ground-based 
$\gamma$-ray astronomy}
\shorttitle{Schwarzschild-Couder telescope for ground-based 
$\gamma$-ray astronomy}
\authors{V. V. Vassiliev and S. J. Fegan.}
\shortauthors{V. V. Vassiliev and S. J. Fegan}
\afiliations{Department of Physics and Astronomy, University of California, Los Angeles, CA 90095, USA}
\email{vvv@astro.ucla.edu}

\abstract{Schwarzschild-type aplanatic telescopes with two aspheric
mirrors, configured to correct spherical and coma aberrations, are
considered for application in $\gamma$-ray astronomy utilizing the
ground-based atmospheric Cherenkov technique. We use analytical
descriptions for the figures of primary and secondary mirrors and, by
means of numerical ray-tracing, we find telescope configurations which
minimize astigmatism and maximize effective light collecting area. It
is shown that unlike the traditional prime-focus Davies-Cotton design,
such telescopes provide a solution for wide field of view $\gamma$-ray
observations. The designs are isochronous, can be optimized to have no
vignetting across the field, and allow for significant reduction of
the plate scale, making them compatible with finely-pixilated cameras,
which can be constructed from modern, cost-effective image sensors
such as multi-anode PMTs, SiPMs, or image intensifiers.}

\begin{document}
\maketitle

\section{Introduction}

All present-day atmospheric Cherenkov telescopes (ACTs) have
prime-focus optical systems (OS) with Davies-Cotton or segmented parabolic
primary mirror surfaces. Although these designs have many benefits and
have proved to be reliable in ACT applications, they may be
incompatible with the demanding requirements of next-generation
large-area arrays of ACTs, which are being planned in Europe
(Cherenkov Telescope Array, CTA), and in the U.S. (Advanced Gamma-ray
Imaging System, AGIS). The desire to significantly improve the angular
resolution, simultaneously increase the field of view, and reduce the
focal plane scale of the telescopes, for compatibility with highly
integrated, multi-pixel photon detectors, motivates research of
alternative designs for ACTs. 

\begin{figure*}[t]
\begin{center}
\includegraphics[width=0.35\textwidth]{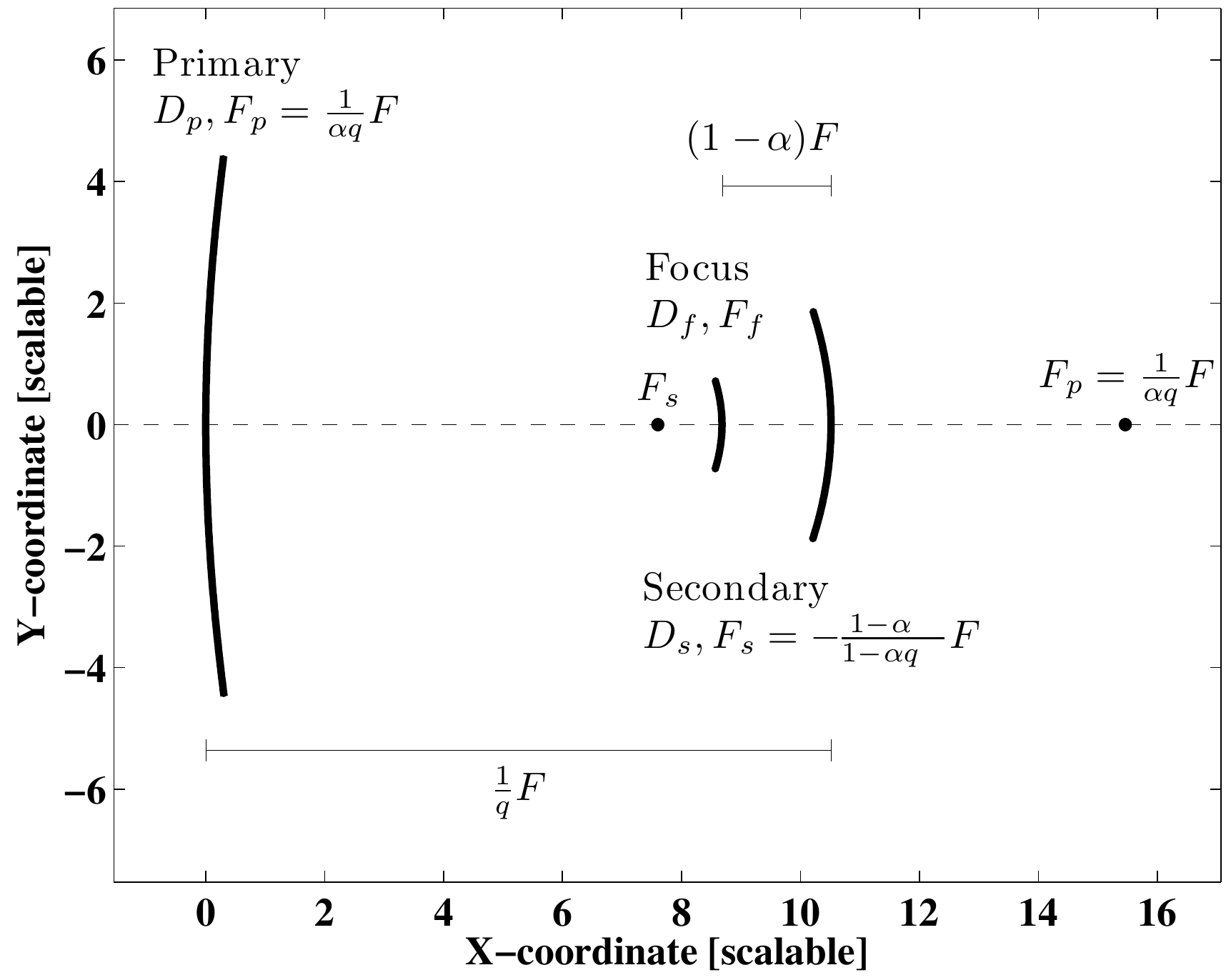}
\end{center}
\caption{Schematic of Schwarzschild-Couder two-mirror optical 
system.}\label{FIG::SYSTEM}
\end{figure*}

Since comatic aberration is the factor limiting the imaging quality of
existing ACTs, we consider two-mirror, aplanatic optical systems which
are free from both coma and spherical aberrations. A subset of these
two-mirror systems with "fast" optics, those with considerably
decreased focal length, are well-suited to the goals of designing
wide-field optics and reducing plate scale. Aplanatic optical systems
were first systematically studied in the classical 1905 paper by Karl
Schwarzschild, in which he described a telescope (shown in
figure~\ref{FIG::SYSTEM}) with two concave mirrors in which the
secondary is placed between the primary and its focus, so that it
de-magnifies the image at the focal plane
\cite{REF::SCHWARZSCHILD::AMKS1905}. This original telescope design
suffered from significant astigmatism, due in part to the additional
requirement of having a flat focal plane that Schwarzschild
imposed. In 1926 Andre Couder derived the mirror surfaces to the third
order in the square of the running radius and showed that astigmatism
could be drastically reduced if the primary and secondary mirrors are
separated by a distance equal to twice the equivalent focal length of
the telescope, and a curved, convex focal plane is introduced
\cite{REF::COUDER::CRAS1926}. The Schwarzschild-Couder (S-C) telescope
has been only infrequently studied since
\cite{REF::WYMAN_KORSCH::AO1975, REF::WILLSTROP::MNRAS1983,
REF::WILLSTROP::MNRAS1984}, and has never been built (although
construction near Paris was initiated by Couder during the 1930s) due
to the sensitivity of its optical performance on the accurate figuring
of the aspheric mirrors and on the exact values of the optical system
parameters, $\alpha$ and $q$ (see definitions on
figure~\ref{FIG::SYSTEM}).

Recently, we attempted to adapt the S-C configuration for wide-field
or small plate scale ACT application
\cite{REF::VASSILIEV::APP2007}. In that study the figures of both
aspheric mirrors were completely constrained by the relative
positioning of the mirrors and the focal plane, and by the
requirements to eliminate on-axis spherical aberrations and to satisfy
the intrinsic symmetry of aplanatic systems given by the ``Abbe sine''
condition. An analytic solution for both aspheric mirrors, in
parametric form, surprisingly exists for all arbitrary two-mirror
optical systems characterized by $(q,\alpha)$, and was first
discovered and described by Linden-Bell in an elegant paper
\cite{REF::LYNDENBELL::MNRAS2002}.

\section{Configurations}

\begin{figure*}[t]
\begin{center}
\includegraphics[width=0.62\textwidth]{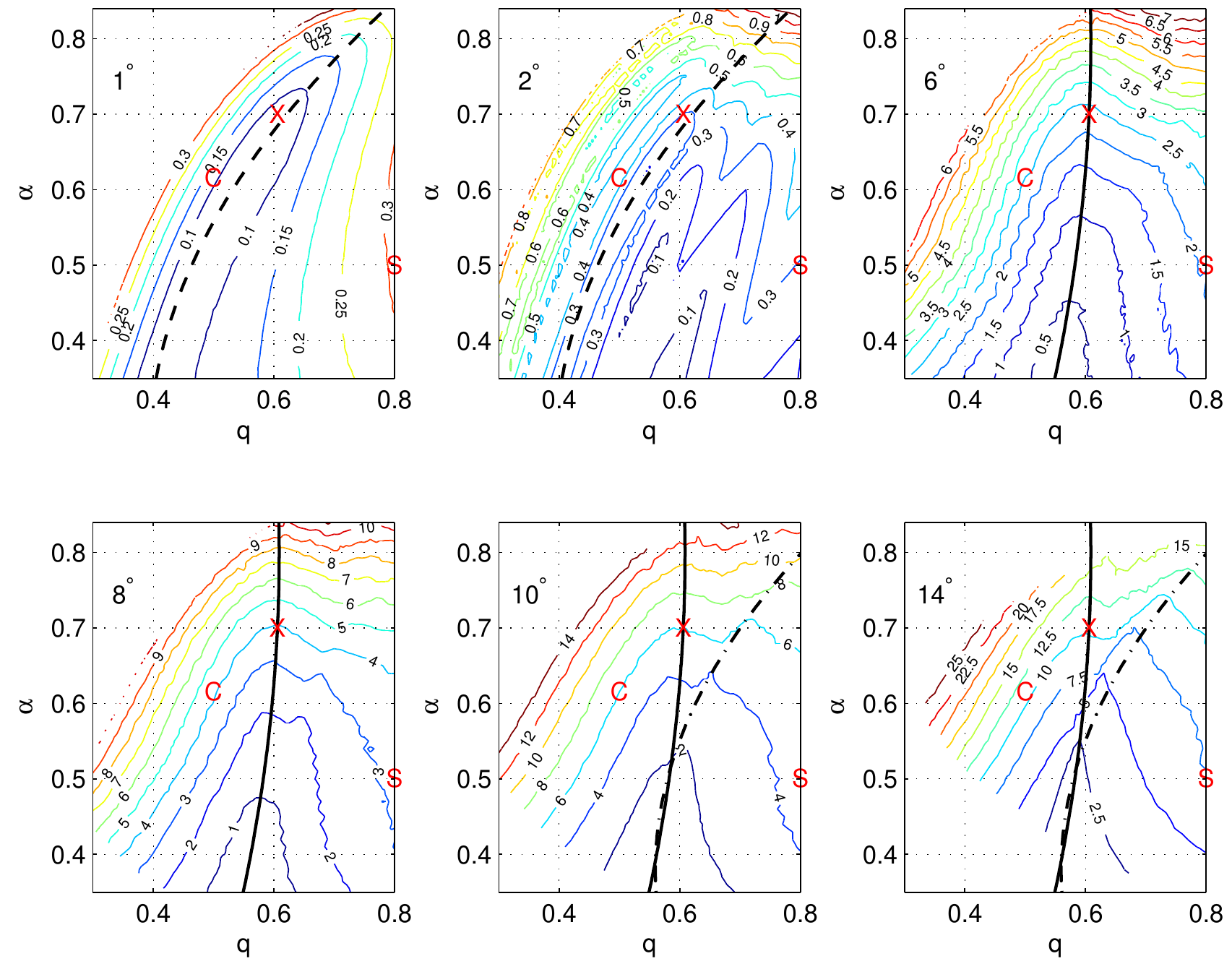}
\end{center}
\caption{$\mathrm{PSF}=2\times \max\{\mathrm{RMS}_{\mathrm{sagittal}},
\mathrm{RMS}_{\mathrm{tangential}}\}$,
as a function of parameters $\alpha$ and $q$ for
systems with different fields of view.}\label{FIG::PSF}
\end{figure*}

Figure~\ref{FIG::PSF} shows the PSF as a function of parameters
$\alpha$ and $q$ for systems with 1, 2, 6, 8, 10, and 14 degrees
un-vignetted fields of view. The lines show OSs with minimal
astigmatism for a given value of $\alpha$. The dashed line indicates
configurations with unvignetted FoV less than $2^\circ$; the Couder
telescope is shown with symbol ``C''. Optimal OSs for FoV in the range
$2^\circ$ to $\sim12^\circ$ are shown with a solid line; symbol ``X''
denotes our choice for ACT applications. The dot-dashed line
illustrates OSs with minimal astigmatism for which the unvignetted FoV
exceeds $12^\circ$; the original solution proposed by Schwarzschild is
denoted as ``S''.

\begin{figure*}[t]
\begin{center}
\includegraphics[width=0.62\textwidth]{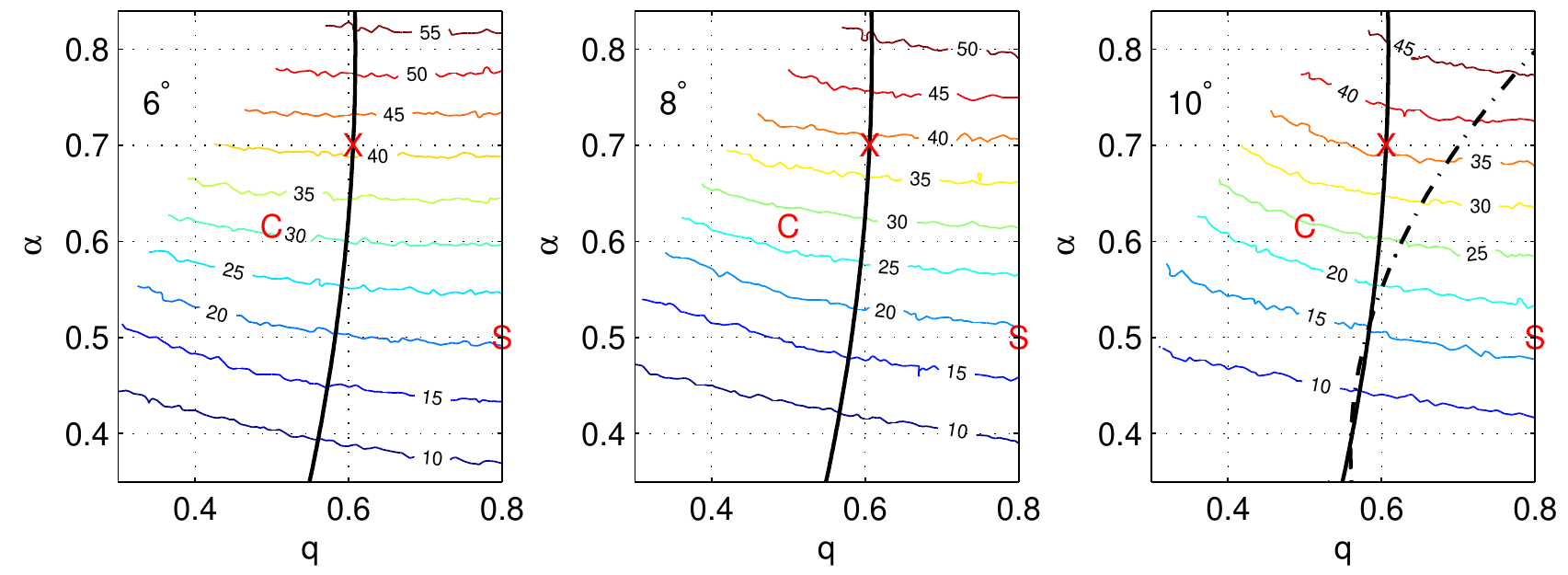}
\end{center}
\caption{Effective light collection area as a function of parameters 
$\alpha$ and $q$ for systems with different fields of view.}\label{FIG::A}
\end{figure*}

\begin{figure*}[t]
\begin{center}
\includegraphics[width=0.62\textwidth]{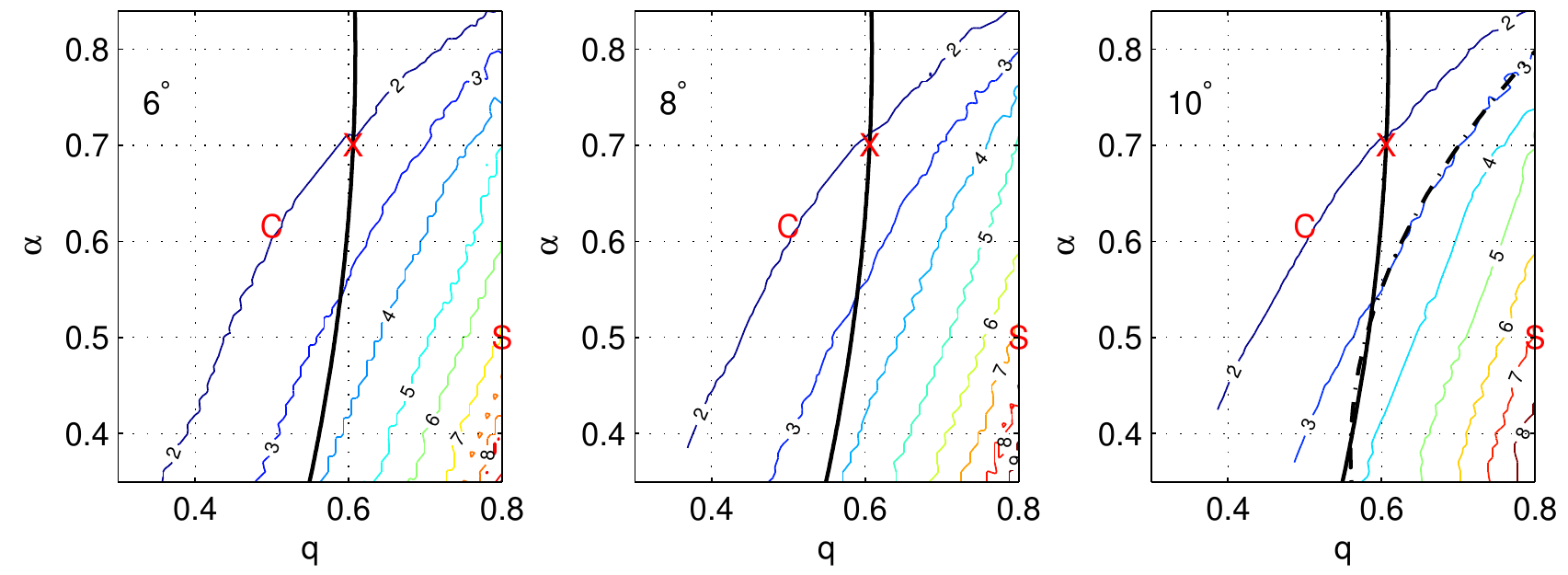}
\end{center}
\caption{Focal plane curvature as a function of parameters 
$\alpha$ and $q$ for systems with different fields of
view.}\label{FIG::F}
\end{figure*}

Figures~\ref{FIG::A} and \ref{FIG::F} show the effective telescope
light collecting area and curvature of the focal plane as a
function of parameters $\alpha$ and $q$ for 6, 8, and 10 degrees
unvignetted FoV. The solutions with minimal astigmatism are indicated as
lines. The focal length of all OSs was chosen to be 500 cm. The
Schwarzschild solution, (``S''), was originally proposed as the design
with minimum curvature of the focal plane. It is evident that both the
telescope effective light collecting area, and the radius of curvature
of the focal plane are effectively one dimensional functions of the
combination of $\alpha$ and $q$ parameters, since the contour isolines
are almost parallel. The light collecting area and maximal incident
angle of photons onto the focal plane are determined predominately by
the value of $\alpha$.

\begin{figure*}[t]
\begin{center}
\includegraphics[width=0.62\textwidth]{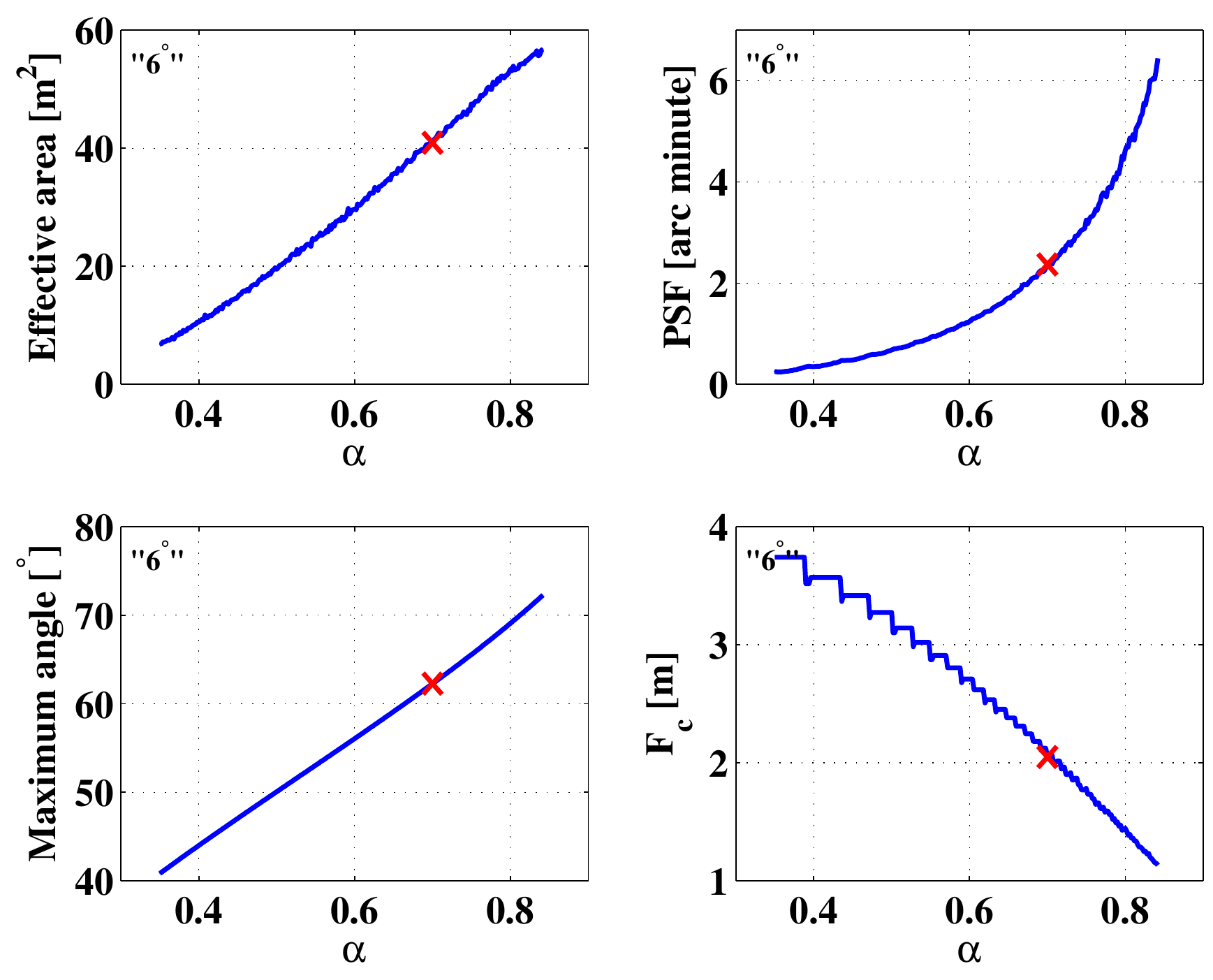}
\end{center}
\caption{The characteristics of OSs optimized for use in ACTs.}
\label{FIG::SIX_DEGREE}
\end{figure*}

The characteristics of OSs optimized for use in ACTs, and which have
no vignetting within the central $6^\circ$ of their FoV, are shown in
figure~\ref{FIG::SIX_DEGREE}. All systems have focal length of F=500
cm, which can be rescaled as required, for example to give a certain
light collecting area, camera plate scale, or maximal angle of
incidence of rays onto the focal plane. The PSF at a field angle of
$3^\circ$ and curvature of the focal plane are shown on the right
figures, as a function of $\alpha$. 

\begin{figure*}[t]
\begin{center}
\resizebox{0.62\textwidth}{!}{\includegraphics[height=0.5\textwidth]{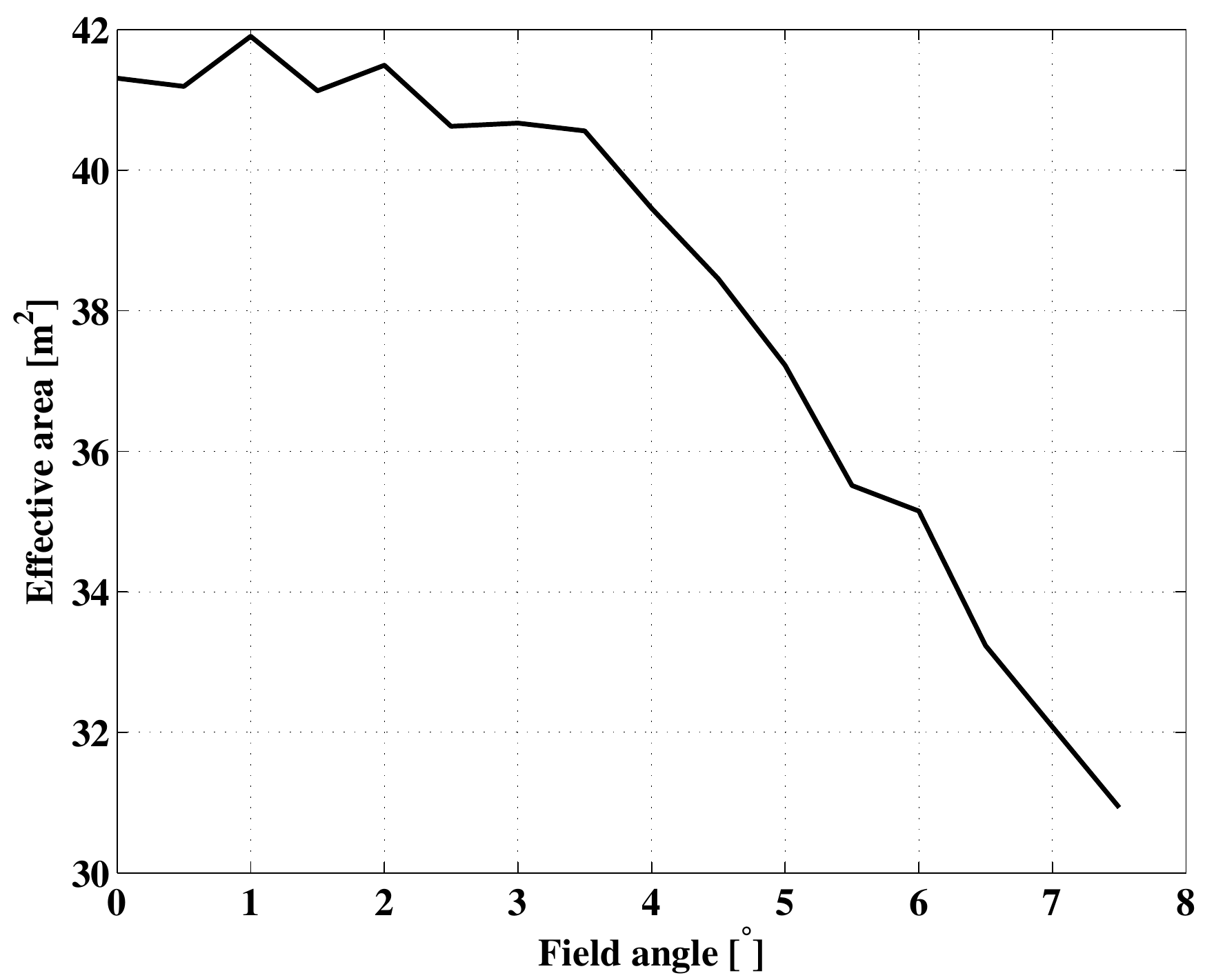}%
\includegraphics[height=0.5\textwidth]{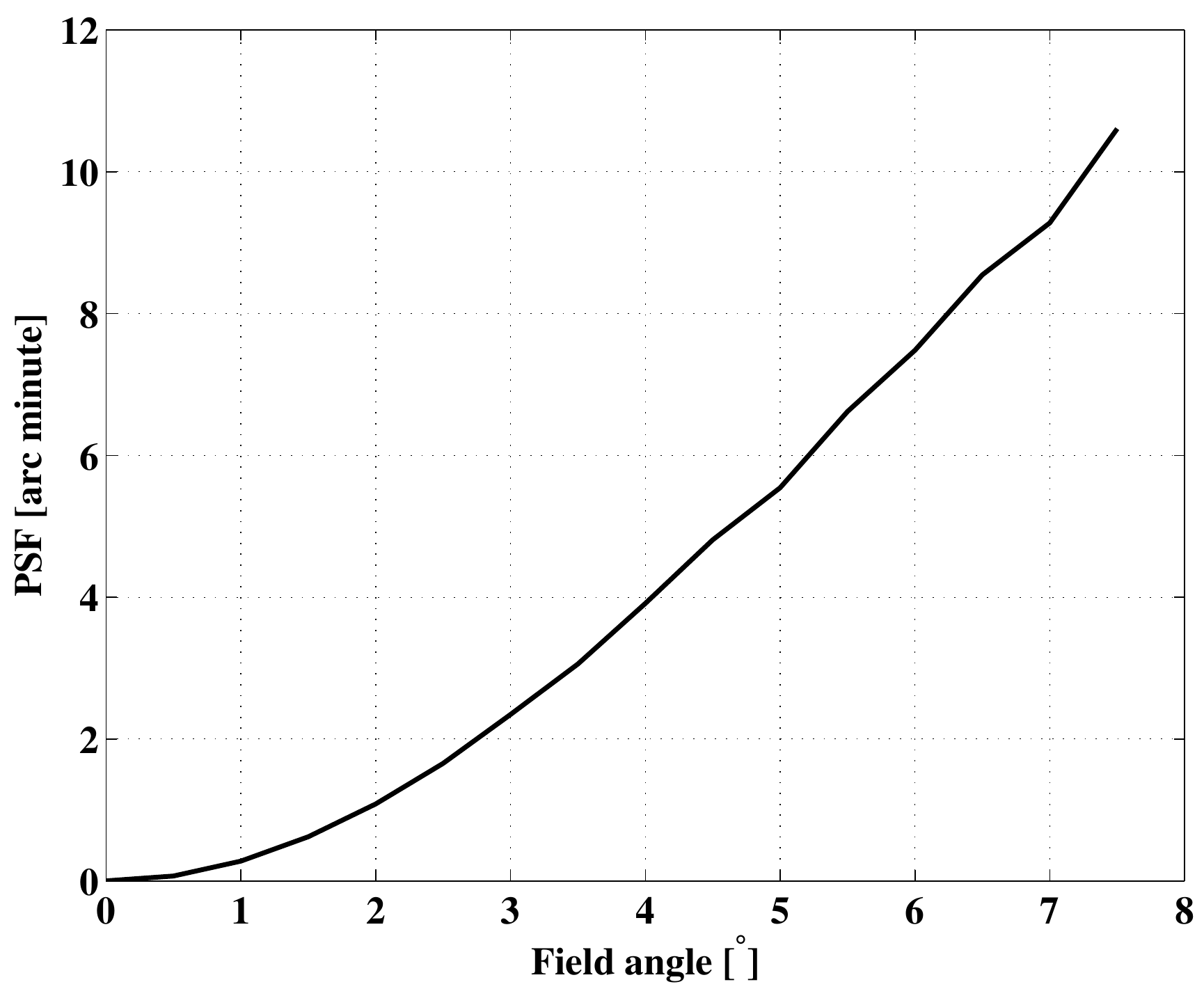}}
\end{center}
\caption{Effective collecting area and PSF for our preferred OS.}
\label{FIG::VS_ANGLE}
\end{figure*}

The effective collecting area and PSF for our preferred OS (with
$\alpha=0.7$, $q=0.606$, which is marked on all figures with the
symbol ``X'', are shown in figure~\ref{FIG::VS_ANGLE} as a function of
field angle.

\section{Conclusions}

Schwarzschild-Couder aplanatic optical system provides a viable
solution for ACTs designed to have small plate scale or to conduct
wide field of view $\gamma$-ray observations. The designs are
isochronous and can be optimized to have no vignetting across the
field. The significantly reduced plate scale makes them compatible
with finely-pixilated cameras, which can be constructed from modern,
potentially cost-effective image sensors such as multi-anode PMTs,
SiPMs, or image intensifiers.

\bibliography{icrc0782}

\begin{thebibliography}{1}

\bibitem{REF::COUDER::CRAS1926}
A.~{Couder}.
\newblock {\em Compt. Rend. Acad. Sci., Paris}, 1276:45--, 1926.

\bibitem{REF::LYNDENBELL::MNRAS2002}
D.~{Lynden-Bell}.
\newblock {\em MNRAS}, 334:787--796, 2002.

\bibitem{REF::SCHWARZSCHILD::AMKS1905}
K.~{Schwarzschild}.
\newblock {\em Astronomische Mittheilungen von der Koeniglichen Sternwarte zu
  Goettingen}, 10:3--28, 1905.

\bibitem{REF::VASSILIEV::APP2007}
V.~{Vassiliev}, S.~{Fegan}, and P.~{Brousseau}.
\newblock {\em \app}, {in press}:{astro--ph/0612718}, 2007.

\bibitem{REF::WILLSTROP::MNRAS1983}
R.V. {Willstrop}.
\newblock {\em MNRAS}, 204:99--103, 1983.

\bibitem{REF::WILLSTROP::MNRAS1984}
R.V. {Willstrop}.
\newblock {\em MNRAS}, 209:587--606, 1984.

\bibitem{REF::WYMAN_KORSCH::AO1975}
C.L. {Wyman} and D.~{Korsch}.
\newblock {\em Applied Optics}, 14:992--995, 1975.

\end{thebibliography}
%This in the bibtex style, is ok.
\bibliographystyle{plain}

\end{document}